\documentclass[pra,twocolumn,showpacs,preprintnumbers,amsmath,amssymb,tightenlines,epsfig,floatfix]{revtex4}

\usepackage{graphicx}
\usepackage{dcolumn}
\usepackage{bm}
\usepackage{amsfonts}

\begin{document}
\preprint{Version: 7.0 \today}
\title{A multi-mode model of a non-classical atom laser produced by outcoupling from a Bose-Einstein condensate with squeezed light}
\author{S.A. Haine and J.J. Hope}
\affiliation{Australian Centre for  Quantum-Atom Optics, Department of Physics, \\
Australian National University, Canberra ACT 0200, Australia}

\pacs{03.75.Pp, 03.70.+k, 42.50.-p}
\begin{abstract}
We examine the properties of an atom laser produced by outcoupling from a Bose-Einstein condensate with squeezed light. We introduce a method which allows us to model the full multimode dynamics of the squeezed optical field and the outcoupled atoms. We show that for experimentally reasonable parameters that the quantum statistics of the optical field are almost completely transferred to the outcoupled atoms, and investigate the robustness to the coupling strength and the two-photon detuning. 
\end{abstract}

\maketitle

\section{Introduction}
Certain precision measurements are improved by using slow-moving massive particles \cite{Rolston}. In a Sagnac interferometer, for example, the inherent sensitivity of a matterwave gyroscope exceeds that of a photon gyoscope with the same particle flux and area by 11 orders of magnitude \cite{Gustavson}.  The high spectral flux and associated first order coherence of atom lasers make them an obvious choice for the input of such devices.  Although current atom laser experiments usually operate in a regime limited by technical noise, the fundamental limit of these measurements will be caused by the shot noise of the atomic field, which will be intrinsic to all interferometers without a non-classical atomic source. Sensitivity is increased in optical interferometry by `squeezing' the quantum state of the optical field, where the quantum fluctuations in one quadrature are reduced compared to a coherent state, while the fluctuations in the conjugate quadrature are increased. In the context of atom optics, it is therefore interesting to ask whether highly squeezed atom optical sources can be produced.

Generation of squeezed atomic beams has been proposed by either utilising the nonlinear atomic interactions to create correlated pairs of atoms via either molecular down conversions or spin exchange collisions \cite{Duan1}, \cite{Pu}, \cite{Drummond}, or by transferring the quantum state of a squeezed optical field to the atomic beam \cite{Moore}, \cite{Jing}, \cite{Fleischhauer}, \cite{Haine1}. 

The generation of nonclassical light is well established experimentally \cite{Bachor}. This suggests that a nonclassical atom laser output could be generated by transferring the quantum state of an optical mode to an atomic beam. Moore {\it et al.} showed that a quantized probe field could be partially transferred to the momentum `side modes' of a condensate consisting of three-level atoms in the presence of a strong pump field, via a Raman transition \cite{Moore}. Jing {\it et al.} performed a single mode analysis of the atom laser outcoupling process for a two-level atom interacting with a quantized light field, and showed that the squeezing would oscillate between the light field and the atomic field at the Rabi frequency \cite{Jing}. Fleischhauer {\it et al.} \cite{Fleischhauer} showed that Raman adiabatic transfer can be used to transfer the quantum statistics of a propagating light field to a continuously propagating beam of atoms by creating a polariton with a spatially dependent mixing angle, such that the output contained the state of the probe beam.

It has been demonstrated \cite{Robins} that the complicated multimode dynamics in atom laser outcoupling can cause effects such as back coupling and a `bound state' significantly limiting the flux.  Although it is well established that a system of three level atoms interacting with a quantized probe field via a Raman transition can exhibit some degree of quantum state transfer, what remains to be demonstrated is whether the multimode effects inherent in atom laser outcoupling will inhibit the clean transfer of the quantum state to the atom laser beam.  In a previous paper \cite{Haine1} we showed that even when we assume that both the probe field and condensate mode are single mode, the outcoupled atoms can still exhibit complicated multimode dynamics. For complete quantum state transfer, the the atoms will have to undergo a quarter Rabi oscillation in the time taken to leave the coupling region, and the light from the probe field will have to be completely absorbed. In our previous paper we assumed a single mode probe field, which is a valid approximation if the light makes many passes through the condensate before it is completely absorbed, as in a high finesse cavity. For a more realistic single pass experiment, a multimode model of the light field is required. In this paper we develop a technique which allows us to model the full multi mode dynamics of the probe beam and the outcoupled atoms, and allows us to investigate how multimode dynamics affect the quantum state transfer.

\section{Model}
\begin{figure}
\includegraphics[width=\columnwidth]{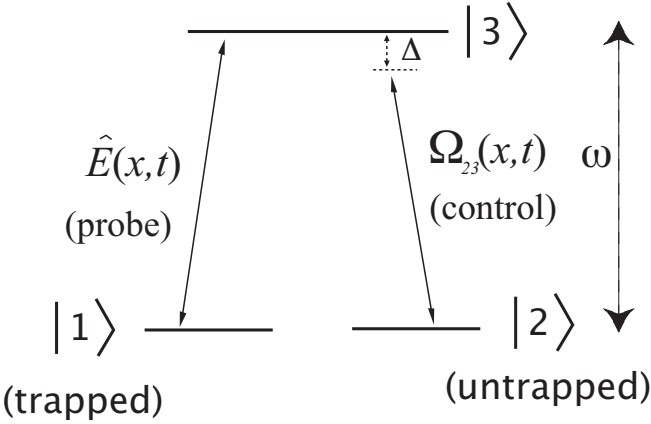}
\caption{\label{fig:levels} Internal energy levels of a three level atom. A condensate of state $|1\rangle$ atoms confined in a trapping potential is coupled to free space via a Raman transition.  The two fields of the Raman transition are a probe beam (annihilation operator $\hat{E}(x,t)$) and a semiclassical control field ($\Omega_{23}(x,t)$) that is detuned from the excited state by an amount $\Delta$.}
\end{figure}

We model an atom laser in one dimension as a BEC of three-level atoms coupled to free space via a Raman transition, as shown in Fig. \ref{fig:levels}. The optical field affecting the $|2\rangle \rightarrow |3\rangle$ transition (pump field) is assumed to be strong and is well approximated by a monochromatic classical field $\Omega_{23}(x,t) = \Omega_{23}e^{i(k_0 x -(\omega_0-\Delta)t)}$ where $\Omega_{23}$ is the single photon Rabi frequency and $\Delta$ is the detuning from the excited state. 

The Hamiltonian for the system is
\begin{eqnarray}
\mathcal{H} &=& \mathcal{H}_{atom} + \mathcal{H}_{int}  + \mathcal{H}_{light} \\ \nonumber
&=& \int \hat{\psi}^{\dag}_1(x) H_0 \hat{\psi}_1(x) dx + \int \hat{\psi}^{\dag}_2(x)(-\frac{\hbar^2}{2m}\nabla^2)  \hat{\psi}_2(x) dx \\ \nonumber
 &+& \int \hat{\psi}^{\dag}_3(x)(-\frac{\hbar^2}{2m}\nabla^2 + \hbar\omega_0) \hat{\psi}_3(x) dx \\ \nonumber
&+& \hbar\int(\hat{\psi}_2(x)\hat{\psi}^{\dag}_3(x)\Omega(x,t) + h.c.)dx \\ \nonumber
&+& \hbar g_{13}\int(\hat{E}(x)\hat{\psi}_1(x)\hat{\psi}^{\dag}_3 + h.c.) dx  + \mathcal{H}_{light} 
\end{eqnarray}
where $\Omega_{23}(x,t) = \Omega_{23} e^{i(k_0 x - (\omega_0 - \Delta)t)}$ where $\Omega_{23}$ is the rabi frequency for the $|2\rangle \rightarrow |3\rangle$ transition, $H_0$ is the single particle Hamiltonian for the trapped atoms, and $m$ is the mass of the atoms, $\hat{\psi}_{1}(x)$ is the annihilation operator for the condensate mode (internal state $|1\rangle$), $\hat{\psi}_3(x)$ is the annihilation operator for the atoms in the excited atomic state ($|3\rangle$), and $\hat{\psi}_2(x)$ is the annihilation operator for the untrapped, freely propagating atoms ($|2\rangle$). The atomic field operators satisfy the usual bosonic commutation relations:
\begin{equation}
[\hat{\psi}_i(x), \quad \hat{\psi}_j(x^{\prime})] =[\hat{\psi}^{\dagger}_i(x), \quad \hat{\psi}^{\dagger}_j(x^{\prime})] =0,\nonumber
\end{equation}
\begin{equation}
[\hat{\psi}_i(x), \quad \hat{\psi}^{\dagger}_j(x')] = \delta_{ij}\delta(x-x')
\end{equation}
$\hat{E}(x)$ is the spatially dependent annihilation operator for the probe field, satisfying
\begin{equation}
[\hat{E}(x), \quad \hat{E}^{\dagger}(x')] = \delta(x-x')
\end{equation}
The coupling coefficient, $g_{13} = \frac{d_{13}}{\hbar}\sqrt{\frac{\hbar\omega_k}{2\epsilon_0}}$, where $d_{13}$ is the dipole moment for the $|1\rangle \rightarrow |3\rangle$ transition. We have assumed that $g_{13}(\omega_k)$ is  approximately flat in the range of interest of our system.

The equations of motion for the Heisenberg operators are:
\begin{eqnarray}
i\dot{\hat{\psi}}_1(x) &=& \frac{H_0}{\hbar} \hat{\psi}_1(x) + g_{13}\tilde{\psi}_3(x)\tilde{E}^{\dagger}(x)  \nonumber \\
i\dot{\hat{\psi}}_2(x) &=& -\frac{\hbar}{2m}\nabla^2\hat{\psi}_2(x) + \Omega_{13}^{*}e^{-ik_0x}\hat{\psi}_3(x) \nonumber \\
i\dot{\tilde{\psi}}_3(x) &=& (-\frac{\hbar}{2m}\nabla^2 + \Delta)\tilde{\psi}_3(x) + \Omega_{23}e^{ik_0x}\hat{\psi}_2(x) \nonumber \\ 
&+& g_{13}\tilde{E}(x)\hat{\psi}_1(x) \nonumber \\
i\dot{\tilde{E}}(x) &=& (-ic\frac{\partial}{\partial x}-(\omega_0 -\Delta))\tilde{E}(x) + g_{13}\hat{\psi}_1(x)\tilde{\psi}_3^{\dagger}(x) \nonumber \\
\end{eqnarray}
where $\tilde{\psi}_3(x) = \hat{\psi}_3e^{i(\omega_0-\Delta)t}$ and $\tilde{E}(x) = \hat{E}e^{i(\omega_0-\Delta)t}$. The population of the excited state $|3\rangle$ will be much less than the other levels when the detunings of the optical fields are larger than the other terms in the system (including the kinetic energy of the excited state). Furthermore, most of the dynamics will occur on time-scales less than $\frac{1}{\Delta}$, so in this regime we can set $\tilde{\psi}_3(x) = -\frac{\Omega_{23}}{\Delta}e^{ik_0x}\hat{\psi}_2(x) - \frac{g_{13}}{\Delta}\tilde{E}(x)\hat{\psi}_1$. If the condensate is approximated by a coherent state containing a large number of atoms and the outcoupling is weak (ie. the condensate wave function is not strongly perturbed by the outcoupling) we can write $\hat{\psi}_1(x,t) = \sqrt{N}\phi_0(x)e^{-i\omega_t t}$, where $\phi_0(x)$ is the ground state wave function of a harmonic trap with trapping frequency $\omega_t$, and $N$ is the mean number of atoms in the condensate. The assumption of ignoring the backaction of the outcoupling on the condensate wavefunction is only valid if the number of photons in the probe beam is much less than the number of atoms in the condensate, as it will need to be in a successful quantum state transfer experiment. We have ignored the atom-atom interactions in our model, which will only be valid if the condensate is dilute. Strong atom-atom interaction will cause complicated dynamics of the quantum state of the condensate mode, which may not necessarily inhibit the efficient generation of a squeezed atom laser, but we can not model this evolution with the techniques used in this paper. Inclusion of such effects would require a more complicated technique, such as a phase space method \cite{phasespace}. With these approximations, our equations of motion for the outcoupled atomic field and the probe field become
\begin{eqnarray}
i\dot{\hat{\psi}}(x) &=& H_a\hat{\psi}(x) - \Omega_C(x)e^{-ik_0x}\tilde{E}(x) \label{psi_eom} \\ 
i\dot{\tilde{E}}(x) &=& H_b\tilde{E}(x) - \Omega_C^{*}(x)e^{ik_0x}\hat{\psi}(x) \label{E_eom}
\end{eqnarray}
with $H_a = (-\frac{\hbar}{2m}\frac{\partial^2}{\partial x^2} -\frac{|\Omega_0|^2}{\Delta} - \omega_t)$, $H_b = (-ic\frac{\partial}{\partial x}- \frac{|g|^2N}{\Delta}|\phi_0(x)|^2 -(\omega_0 -\Delta))$,  $\Omega_C(x) = \frac{\Omega^{*}_{23}g_{13}}{\Delta}\sqrt{N}\phi_0(x)$, and $\hat{\psi} = \hat{\psi}_2 e^{i\omega_t t}$. 

Equations of the form of (\ref{psi_eom}) and (\ref{E_eom}) are common in quantum atom optics problems. In the next section we introduce our technique for solving these equations.

\section{Solution method}
At $t=0$ the field operators can be expanded as
\begin{eqnarray}
\hat{\psi}(x, t=0) &=& \sum_i f_i(x) \hat{a}_i \\
\tilde{E}(x, t=0) &=& \sum_i p_i(x) \hat{b}_i 
\end{eqnarray}
where $f_i(x)$ and $p_i(x)$ represent an expansion in any orthonormal basis, the operators $\hat{a}_i$ and $\hat{b}_i$ represent Schr\"{o}dinger picture annihilation operators for the $i$th mode of the atomic and optical fields respectively, and $\hat{a}^{\dagger}_i\hat{a}_i$ ($\hat{b}^{\dagger}_i\hat{b}_i$) represent the number of atoms (photons) in mode $i$. From this we can postulate that the solution to Eq. (\ref{psi_eom}) and Eq. (\ref{E_eom}) is
\begin{eqnarray}
\hat{\psi}(x, t) &=& \sum_i f_i(x, t)\hat{a}_i + \sum_i g_i(x,t) \hat{b}_i \label{psi_sol} \\
\tilde{E}(x, t) &=& \sum_i p_i(x, t)\hat{b}_i + \sum_i q_i(x,t) \hat{a}_i \label{Esol}
\end{eqnarray}
By substituting this ansatz into Eq. (\ref{psi_eom}) and Eq. (\ref{E_eom}) we obtain equations of motion for the mode functions $f_i(x)$, $g_i(x)$, $p_i(x)$ and $q_i(x)$ is given by
\begin{eqnarray}
i\dot{f}_i(x) &=& H_a f_i(x) - \Omega_c e^{-ik_0x} q_i(x) \\
i\dot{g}_i(x) &=& H_a g_i(x) - \Omega_c e^{-ik_0x} p_i(x)  \label{geom} \\ 
i\dot{p}_i(x) &=& H_b p_i(x) - \Omega_c^{*}e^{ik_0x} g_i(x) \label{peom} \\
i\dot{q}_i(x) &=& H_b q_i(x) - \Omega_c^{*}e^{ik_0x} f_i(x) 
\end{eqnarray} 
with $g_i(x, t=0)=q_i(x, t=0)=0$. In practice we are free to choose any initial conditions we like for the $f_i(x)$s and $p_i(x)$s, as long as they form an orthonormal basis. From the solutions to these equations we can obtain the dynamics of any observable of the system. However, keeping track of all the mode functions (particularly when a numerical solution is required) can be daunting. In our system we initially have no outcoupled atoms, and if we choose our basis carefully, we can choose it so that the photons only initially occupy one mode. i.e. $|\Psi(t=0)\rangle = |0, 0, ..0\rangle_{atoms}\otimes|\gamma, 0, ...,0\rangle_{light}$ where $|\gamma\rangle$ represents an arbitrary state of a single optical mode.  In our subsequent calculations we choose the mode basis to be plane waves, with $\hat{b}_0$ representing the annihilation operator for a plane wave with momentum $\mathbf{k}_p$. This means that the initial condition of $p_0(x)$ is a plane wave with momentum $\mathbf{k}_p$. 

By noticing that $\hat{a}_i$, for all $i$ and $\hat{b}_i$, $i\neq0$ acting on our state return zero for all time, we can write Eq. (\ref{psi_sol}), and Eq. (\ref{Esol}) in a more illuminating way.
\begin{eqnarray}
\hat{\psi}(x,t) &=& g_0(x,t)\hat{b}_0 + \hat{V}_{\psi}(x,t) \\
\tilde{E}(x,t) &=& p_0(x, t)\hat{b}_0 + \hat{V}_E(x,t)
\end{eqnarray}
with
\begin{eqnarray}
\hat{V}_{\psi}(x, t) &=& \sum_i f_i(x, t) \hat{a}_i + \sum_{i\neq0}g_i(x,t)\hat{b}_i \\
\hat{V}_{E}(x, t) &=&  \sum_i q_i(x, t) \hat{a}_i + \sum_{i\neq0}p_i(x,t)\hat{b}_i
\end{eqnarray}

Instead of solving for the coefficients of the annihilation operators in $\hat{V}_{\psi}$ and $\hat{V}_E$ individually, we note that since we know how they operate on our state for all time, the only extra information we need are their commutation relations.  By noticing that $[\hat{\psi}(x,t), \quad \hat{\psi}^{\dag}(x',t)] = [\hat{E}(x,t), \quad \hat{E}^{\dag}(x',t)] = \delta(x-x')$ and $[\hat{a}_i, \quad \hat{a}^{\dag}_j] =  [\hat{b}_i, \quad \hat{b}^{\dag}_j] = \delta_{ij}$ with all other commutators zero, it is easy to show that 
\begin{equation}
[\hat{V}_{\psi}(x,t), \hat{V}_{\psi}^{\dag}(x', t)] = \delta(x-x') - g_0^{*}(x',t)g_0(x,t) 
\end{equation}
\begin{equation}
[\hat{V}_E(x,t),  \hat{V}_E^{\dag}(x', t)] = \delta(x-x') - p_0^{*}(x',t)p_0(x,t) \\
\end{equation}

From the solutions for $g_0(x,t)$, $p_0(x, t)$ and these commutation relations, we can calculate any observable of the system. This method converts the full multi-mode quantum field system into two coupled classical modes that change dynamically with time. 

To reduce computational burden when solving for $g_0(x,t)$ and $p_0(x,t)$ numerically, we have transformed to the momentum shifted variables $\tilde{p}_0(x) = p_0(x)e^{-ik_px}$ and $\tilde{g}_0(x) = g_0(x)e^{i(\mathbf{k_0}-\mathbf{k_p})x}$ giving us
\begin{eqnarray}
i\dot{\tilde{g}}_0(x) &=& (-\frac{\hbar}{2m}\nabla^2 - \frac{\hbar}{m}|\mathbf{k}_0 -\mathbf{k}_p|\nabla 
 \label{gtilde_eom} \\
&+& \frac{\hbar}{2m}|\mathbf{k}_0
 - \mathbf{k}_p|^2 - \frac{|\Omega_{23}|^2}{\Delta} -\omega_t)\tilde{g}_0(x) - \Omega_c\tilde{p}_0(x) \nonumber\\
 i\dot{\tilde{p}}_0(x) &=& (-ic\nabla + \delta -\frac{|g_{13}|^2}{\Delta}N|\phi_0(x)|^2)\tilde{p}_0(x) \nonumber \\
  &-& \Omega^{*}_c(x)\tilde{g}_0(x) 
 \end{eqnarray} 
 with $\delta = c|\mathbf{k_p}|- c|\mathbf{k_0}|$ (since $c|\mathbf{k_0}| = \omega_0 - \Delta$) being the two photon detuning. Physically, $\mathbf{k}_p$ represents the momentum of the initial mode of the optical state with non-zero occupation, as the initial condition on $p_0(x)$ defines the mode that is occupided. Optimum coupling between the atomic and optical fields will occur when they are resonant. This will occur when $\delta \approx \frac{\hbar}{2m}|\mathbf{k}_0-\mathbf{k}_p|^2+\frac{|g_{13}|^2N}{\Delta}|\phi_0(0)|^2 - \frac{|\Omega_{23}|^2}{\Delta} -\omega_t \equiv \delta_0$. 

By noticing that the evolution of the optical field is trivial at high frequencies (of order $c|\mathbf{k}_p|$), and the evolution we are interested in (the interaction with the atomic field) will occur on frequency scales of less than $\Omega_C$, we can make the slow envelope approximation \cite{duttonthesis} and approximate the dynamics of $p_0(x,t)$ as

\begin{equation}
ic\frac{dp_0(x)}{dx} = \big(\delta -\frac{|g|^2N}{\Delta}|\phi_0(x)|^2)p_0(x\big) - \Omega_c^*g_0(x) \label{peom2}
\end{equation}
 
In the following section we will look at the solutions to these equations and use them to look at some properties of the outcoupled atoms. 

\section{Properties of the outcoupled atoms}
\begin{figure}
\includegraphics[width=\columnwidth]{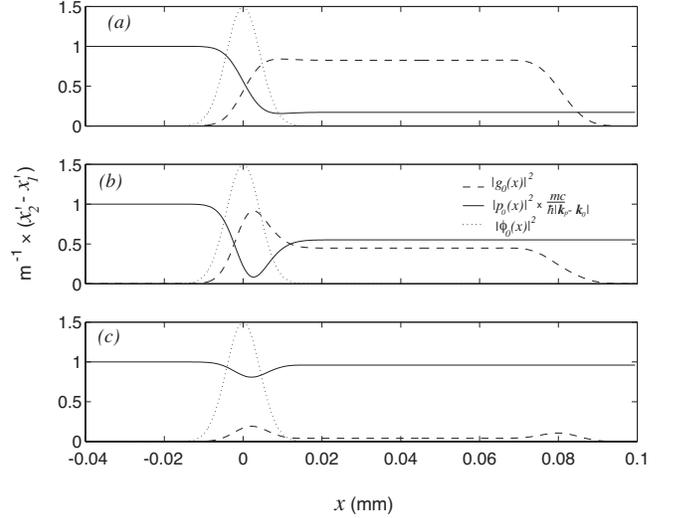}
\caption{\label{fig:densitys} $|g_0(x)|^2$ (dashed line) and $|p_0(x)|^2$ (solid line) at $t=7.2$ ms as found numerically for (a) $\delta = \delta_0$ and $\Omega_{23} = 2.1\times 10^{12}$ rad s$^{-1}$, (b) $\delta = \delta_0$ and $\Omega_{23} = 3.2\times 10^{12}$ rad s$^{-1}$ and (c), $\delta = \delta_0 + 4\times 10^{3}$ rad s$^{-1}$ and $\Omega_{23} = 2.0\times 10^{12}$ rad s$^{-1}$ with all other parameters given in the text. The densities are related to the functions $|g_0(x,t)|^2$ and $|p_0(x,t)|^2$ by a factor of $\langle\hat{b}_0^{\dag}\hat{b}_0\rangle$, ie. the initial optical density of the field. The dotted line  ($|\phi_0(x)|^2$) represents the density profile of the condensate (density in arbitrary units). In (a), the probe beam is attenuated as it passes through the condensate, producing a steady flux of atoms. In (b) the coupling strength is too strong for ideal quantum transfer, causing significant back coupling of atoms into the condensate, which causes a bound state of atoms and reduces the flux. In (c), the two photon detuning is such that the light in the probe beam is off resonant with the Raman transition. An initial pulse of atoms is ejected from the condensate, then the flux reduces as atoms remain bound to the condensate. $p_0(x)$ was normalized at $t=0$ such that $\frac{mc}{\hbar|\mathbf{k}_p-\mathbf{k}_0|}\int_{x_1}^{x_2}|p(x)|^2dx = 1$ with $x_2-x_1 = 0.02$ mm.}
\end{figure}

The density of outcoupled atoms and the optical density (mean number of photons per unit length) are given by
\begin{eqnarray*}
\langle \hat{\psi}^{\dag}(x) \hat{\psi}(x)\rangle = |g(x)|^2\langle \hat{b}_0^{\dag}\hat{b}_0 \rangle  \\
\langle \hat{E}^{\dag}(x) \hat{E}(x)\rangle = |p(x)|^2\langle \hat{b}_0^{\dag}\hat{b}_0 \rangle
\end{eqnarray*}
The observables $\langle \hat{\psi}^{\dag}(x) \hat{\psi}(x)\rangle$ and $\langle \hat{E}^{\dag}(x) \hat{E}(x)\rangle$ are only related to our solutions for $g_0(x,t)$ and $p_0(x,t)$ by such a simple expression because of the specific choice of the basis for the optical field. We solved Eq. (\ref{gtilde_eom}) and Eq. (\ref{peom2}) numerically for $g_0(x,t)$ and $p_0(x, t)$ using a 4th order Runge Kutta algorithm with a cross propagation step using the numerical package XMDS \cite{xmds}. We chose parameters realistic to atom optics experiments with Rb$^{87}$ atoms. Unless stated otherwise, we have set $m = 1.4 \times 10^{-25}$ kg, $\omega_t = 20.0$ rad s$^{-1}$, $g_{13} = 28.9$ rad s$^{-1}$, $N = 10^6$ and $\Delta = 10^{11}$ rad s$^{-1}$.   Fig. \ref{fig:densitys} shows typical results for the densities $|g_0(x)|^2$ and $|p_0(x)|^2$ after $t=7.2$ ms of evolution.

For optimum transfer of squeezing from the optical field to the atomic field, the quantum efficiency of the outcoupling process (i.e. number of atoms per photon in the probe beam) will have to approach one. When this is the case, the probe field will be completely absorbed. Fig. \ref{fig:densitys} shows that overcoupling can significantly reduce this efficiency. 

\begin{figure}
\includegraphics[width=\columnwidth]{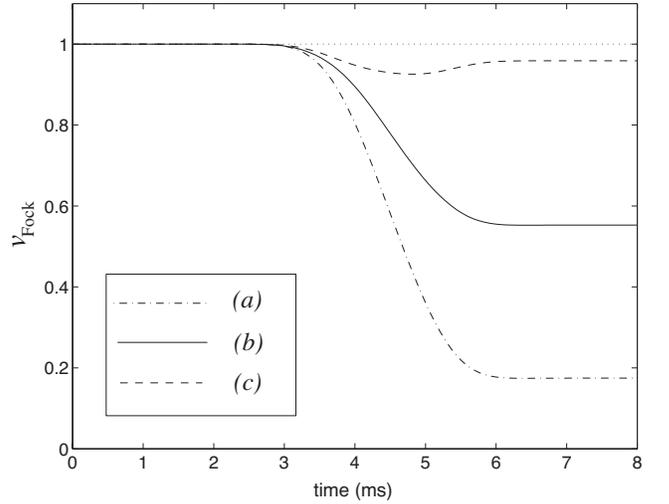}
\caption{\label{fig:numbervar1} $v_{Fock}$ versus time for (a) $\delta = \delta_0$ and $\Omega_{23} = 2.1\times 10^{12}$ rad s$^{-1}$, (b) $\delta = \delta_0$ and $\Omega_{23} = 3.2\times 10^{12}$ rad s$^{-1}$, and (c) $\delta = \delta_0 + 4\times 10^3$ rad s$^{-1}$ and $\Omega_{23} = 2\times 10^{12}$ rad s$^{-1}$.  As the flux of the atom laser beam becomes steady, the $v_{Fock}$ reaches it's minimum value, which is limited by the quantum efficiency of the outcoupling process. The detector region was chosen such that $x_1 = 0.04$ mm, $x_2 = 0.06$ mm.}
\end{figure}

Reduction in the variance of the flux of an atom laser beam would be a measure of how much squeezing is transferred to the atom laser from the optical field. However, in a fully multimode model, the variance of the flux is infinite. Instead, we use the density integrated over a small region as a measure of how `quiet' the atom laser beam is. We define the operator
\begin{equation}
\hat{N} = \int_{x_1}^{x_2} \hat{\psi}^{\dag}(x) \hat{\psi}(x)dx,
\end{equation}
which represents the number of atoms in a region of space between $x_1$ and $x_2$ in the path of the atom laser beam. Physically this operator would be a measure of the `intensity' of the atom laser beam. The variance of $\hat{N}$ is

\begin{eqnarray*}
V(\hat{N}) &=& \langle \hat{N}^2\rangle - \langle \hat{N}\rangle^2 \\
&=& \int\int\langle\hat{\psi}^{\dagger}(x')\hat{\psi}(x')\hat{\psi}^{\dagger}(x)\hat{\psi}(x) \rangle dx dx' \\
&-& \Big(\int \langle \hat{\psi}^{\dagger}(x)\hat{\psi}(x)\rangle dx\Big)^2 \\
&=&  \int \int \langle \hat{\psi}^{\dag}(x')\hat{\psi}^{\dag}(x)\hat{\psi}(x') \hat{\psi}(x)\rangle dx dx' \nonumber \\
&+&  \int\langle \hat{\psi}^{\dag}(x) \hat{\psi}(x) \rangle dx - \Big( \int\langle \hat{\psi}^{\dag}(x) \hat{\psi}(x) \rangle dx \Big)^2
\end{eqnarray*}
Using our solution for $\hat{\psi}(x,t)$ this becomes

\begin{equation}
V(\hat{N}) = N_g^2\Big(\langle \hat{b}_0^{\dagger}\hat{b}_0^{\dagger} \hat{b}_0\hat{b}_0\rangle -  \langle \hat{b}_0^{\dagger} \hat{b}_0\rangle^2\Big) + N_g\langle \hat{b}_0^{\dagger} \hat{b}_0\rangle 
\end{equation}
where $N_g = \int_{x_1}^{x_2}|g(x)|^2 dx$. To compare this to the initial amount of squeezing in the optical field, we define the operator $\hat{N}_0 = \int_{x_1'}^{x_2'}\hat{E}^{\dagger}(x)\hat{E}(x) dx$, where $x_2' - x_1' = \frac{mc}{\hbar|\mathbf{k}_p-\mathbf{k}_0|}(x_2-x_1)$, ie. the range of integration for the optical fields is larger than for the atomic field by a factor of the ratio of the light speed to the mean atomic speed. This is because if we have a well defined number of photons in a region of length $L$ initially, if there is ideal quantum transfer, the number of particles will now be confined to a region of length $\frac{\hbar |\mathbf{k}_p-\mathbf{k}_0|}{mc}L$.  The variance at $t=0$ is then
\begin{equation}
V(\hat{N}_0) =  N_p^2\Big(\langle \hat{b}_0^{\dagger}\hat{b}_0^{\dagger} \hat{b}_0\hat{b}_0\rangle -  \langle \hat{b}_0^{\dagger} \hat{b}_0\rangle^2\Big) + N_p\langle \hat{b}_0^{\dagger} \hat{b}_0\rangle
\end{equation}
where $N_p = \int_{x_1'}^{x_2'} |p(x,t=0)|^2dx \equiv 1$, since we have normalised $p_0(x,t=0)$ such that $\langle \hat{b}^{\dagger}_0\hat{b}_0\rangle$ represents the number of particles per length $x_2' - x_1'$. Since we have assumed that the initial state of the occupied mode for the optical field is a plane wave, the variance is homogenous along the length of the beam at $t=0$. The normalized variance $v(\hat{N}_0) = \frac{V(\hat{N}_0)}{\langle \hat{N}_0\rangle}$ is then
\begin{equation}
 v(\hat{N}_0) = \frac{\Big(\langle \hat{b}_0^{\dagger}\hat{b}_0^{\dagger} \hat{b}_0\hat{b}_0\rangle -  \langle \hat{b}_0^{\dagger} \hat{b}_0\rangle^2\Big)}{\langle \hat{b}_0^{\dagger} \hat{b}_0\rangle} +1
 \end{equation}
 In terms of $v(\hat{N}_0)$, $v(\hat{N}) = \frac{V(\hat{N})}{\langle \hat{N}\rangle}$ is
 \begin{equation}
 v(\hat{N}) = N_g v(\hat{N_0}) +(1-N_g)
 \end{equation}
 As $N_g\rightarrow 1$, $v(\hat{N}) \rightarrow v(\hat{N_0})$. If the initial state of the optical field had perfectly well defined number (i.e. a Fock state), when $v(\hat{N}_0) = 0$, and $v(\hat{N}) = 1-N_g \equiv v_{Fock}$. If a coherent (or vacuum) state is used to outcouple then $v(\hat{N}_0) = 1$, regardless of the efficiency of the outcoupling process. Fig. [\ref{fig:numbervar1}] shows $v_{Fock}$ versus time for the cases shown in Fig. [\ref{fig:densitys}].  
 
 To get complete quantum state transfer (and hence the minimum possible variance in the flux) the coupling between the optical and atomic field will have to be strong enough such that the optical field is completely absorbed. However, if the coupling is two strong, there will be significant back coupling of the atoms into the condensate as was seem in reference \cite{Robins}, \cite{Haine1}. We can estimate the optimum coupling desired by equating the average time taken for an atom to leave the coupling region (condensate) $T_{leave} = \sqrt{\frac{\hbar}{m\omega_t}}m/(\hbar(\mathbf{k}_p - \mathbf{k}_0))$ where $\sqrt{\frac{\hbar}{m\omega_t}}$ is the spatial width of the condensate, with the quarter period Rabi oscillation $\frac{T_{Rabi}}{4} = \frac{\pi}{2\int\Omega_c(x)dx}$. This will occur when $\Omega_{23} \approx 2.3 \times 10^{12}$ rad s$^{-1}$.  

\begin{figure}
\includegraphics[width=\columnwidth]{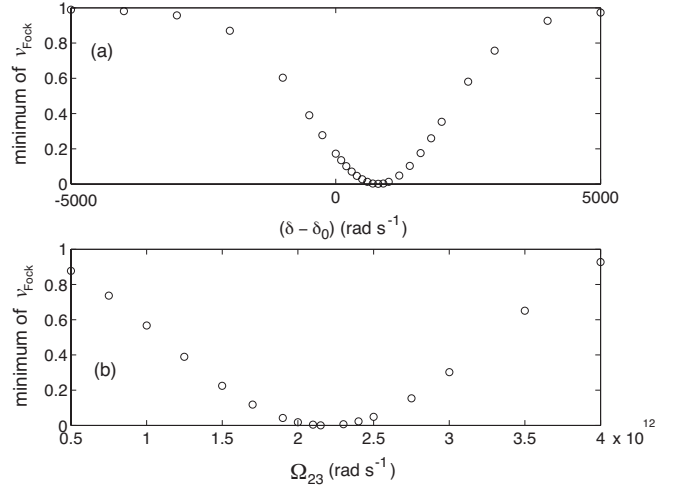}
\caption{\label{fig:deltaplot} (a) The minimum value of $v_{Fock}$ as a function of $\delta - \delta_0$ for $\Omega_{23} = 2.15\times 10^{12}$ rad s$^{-1}$ with all other parameters given in the text. Optimum squeezing occurs at around $\delta-\delta_0 =  800$ rad s$^{-1}$. (b) Minimum value of $v_{Fock}$ as a function of $\Omega_{23}$ for $\delta-\delta_0 = 800$ rad s$^{-1}$. Optimum squeezing occurs at around $\Omega_{23} = 2.2$ rad s$^{-1}$. }
\end{figure}

Obviously, the amount of squeezing that can be transferred to the atom laser beam is strongly dependent on the strength of the coupling and the two-photon de-tuning. These factors introduce complicated multi-mode behaviour into the atom laser beam which reduces the flux and also will broaden the line width, as the atomic mode becomes less monochromatic. Fig. [\ref{fig:deltaplot}] shows the minimum value of $v_{Fock}$ obtained for different values of $\delta$ and $\Omega_{23}$.

Our numerical simulations show that at the appropriate coupling strength and two photon de-tuning, $v_{Fock}$ tends to a very small number $< 0.01$, indicating that the squeezing from optical field is completely transferred to the atom laser beam. However, we cannot give a quantitative limit, as we are limited by round off error in our calculation of $N_g$. Obviously for complete transfer of the squeezing to occur, the light would have to be completely attenuated as it passes through the outcoupling region. Our simulations show that when $\Omega_{23} = 2.2\times 10^{12}$ rad s$^{-1}$, then the squeezing is transferred with an efficiency greater than $0.99$ while the optical field is attenuated by a factor of $10^4$. In any realistic experiment the squeezing measured in the atom laser beam would be limited by other factors such as limited squeezing in the optical beam and detection efficiencies, rather than by this level of quantum state transfer. 

\section{Conclusion}
We have shown that a multimode model of atom laser out-coupling can display a very high level of quantum state transfer when outcoupled under appropriate conditions. This transfer is degraded when the system is overcoupled or off-resonant. The conditions necessary for optimum quantum state transfer are also the once which give the maximum clean flux for the atom laser beam. 

This research was supported by the Australian Research Council.  The Australian Centre for Quantum-Atom Optics is an ARC Centre of Excellence.

\end{document}